\documentclass[onecolumn,floatfix,showpacs,nofootinbib,11pt,groupedaddress,superscriptaddress,]{revtex4-1}
\usepackage{amssymb,amsmath,graphicx}
\usepackage{color}
\usepackage[colorlinks,hyperindex]{hyperref}
\hypersetup{hidelinks,backref=true,pagebackref=true,hyperindex=true,colorlinks=true,breaklinks=true,urlcolor= blue}
\hypersetup{%
  colorlinks = true,
  linkcolor  = blue
}
\definecolor{green1}{RGB}{0,128,0} 
\hypersetup{hidelinks,backref=true,pagebackref=true,hyperindex=true,colorlinks=true,breaklinks=true,urlcolor= blue}
\hypersetup{%
  colorlinks = true,
  linkcolor  = blue,
  citecolor = green1,
}
\usepackage{bookmark}
\usepackage{hyperref,color,xcolor}
\hypersetup{hidelinks,backref=true,pagebackref=true,hyperindex=true,colorlinks=true,breaklinks=true,urlcolor= blue}
\hypersetup{%
  colorlinks = true,
  linkcolor  = blue
}
\usepackage{makeidx,upgreek}
\newcommand{\beq}{\begin{eqnarray}}
\newcommand{\benu}{\begin{enumerate}}
\newcommand{\enu}{\end{enumerate}}
\newcommand{\eeq}{\end{eqnarray}}

\newcommand{\mpl}{m_p}

\renewcommand{\d}{{\rm d}}
\newcommand{\be}{\begin{equation}}
\newcommand{\ee}{\end{equation}}

\newcommand{\ba}{\begin{eqnarray}}
\newcommand{\ea}{\end{eqnarray}}
\newcommand{\bse}{\begin{subequations}}
\newcommand{\ese}{\end{subequations}}
\newcommand{\bea}{\begin{eqnarray}}
\newcommand{\eea}{\end{eqnarray}}

\def\ft#1#2{{\textstyle{{\scriptstyle #1}\over {\scriptstyle #2}}}}

\def\sst#1{{\scriptscriptstyle #1}}

\def\dalemb#1#2{{\vbox{\hrule height .#2pt
        \hbox{\vrule width.#2pt height#1pt \kern#1pt
                \vrule width.#2pt}
        \hrule height.#2pt}}}

 \newcommand{\bltx}{\textcolor{black}}
 
 \def\0{{\sst{(0)}}}
\def\1{{\sst{(1)}}}
\def\2{{\sst{(2)}}}
\def\3{{\sst{(3)}}}
\def\4{{\sst{(4)}}}
\def\5{{\sst{(5)}}}
\def\6{{\sst{(6)}}}
\def\7{{\sst{(7)}}}
\def\8{{\sst{(8)}}}
\def\9{{\sst{(9)}}}

\def\Im{{{\frak{Im}}}}

\begin{document}
\title{GUP black hole remnants in quadratic gravity }
\author{Iber\^e Kuntz}
\email{ibere.kuntz@ufabc.edu.br}
\affiliation{Federal University of ABC, Center of Mathematics,  Santo Andr\'e, 09580-210, Brazil}
\affiliation{Federal University of ABC, Center of Physics,  Santo Andr\'e, 09580-210, Brazil.}
\author{Rold\~ao~da~Rocha}
\email{roldao.rocha@ufabc.edu.br}
\affiliation{Federal University of ABC, Center of Mathematics,  Santo Andr\'e, 09580-210, Brazil}
\affiliation{
International Centre for Theoretical Physics (ICTP), Strada Costiera 11, 34151 Trieste, Italy}
\begin{abstract}
\bltx{The Hawking radiation of static, spherically symmetric, asymptotically flat solutions in quadratic gravity is here scrutinized,  in the context of the generalized uncertainty principle (GUP). Near-center and near-horizon Frobenius expansions of these solutions are studied.
Their Hawking thermal spectrum is investigated out of the tunnelling method and the WKB procedure. Computing the Hawking flux of these black hole solutions shows that, for small black holes and for a precise combination of 
the GUP parameter and the parameters that govern the gravitational interaction in quadratic gravity, the black hole luminosity can vanish. This yields absolutely stable mini black hole remnants in quadratic gravity.}
\end{abstract}
\maketitle
\section{Introduction}
\par
\bltx{Black hole thermodynamics has been occupying a prominent spot in physics in the last five decades, since the  Bekenstein's conjecture was posed, asserting that black hole physics has a close relationship with the laws of thermodynamics \cite{hh1}. Thereafter, Hawking demonstrated that black holes can indeed radiate, when quantum effects set in \cite{hh2,hh3,hh4}. The fact that the black hole radiation is purely thermal 
can state that black holes do have a well-defined temperature, being thermodynamical compact objects. There exist diverse procedures to study the Hawking radiation and the temperature of a black hole. Among them, the tunnelling method is a particularly interesting method for calculating
black hole temperature since it provides a dynamical model of the black hole radiation.}

The Hawking evaporation consists of a quantum effect involving black holes, irrespectively of their masses.
It is usually described via the tunnelling procedure~\cite{2,Vanzo:2011wq,k0,k1} in the WKB semiclassical approximation.
 In the fermionic sector, the Hawking radiation spectrum
 was studied as the tunnelling of fermions satisfying the Dirac equation through an event horizon.
 The tunnelling method has also been used  to calculate the Hawking flux of dark
fermions~\cite{hoff,daRocha:2005ti} across the event horizon of black hole geometries. This method also encompasses small black holes, whose masses are of the order of the Planck scale \cite{Casadio:2017sze,daRocha:2006ei,Casadio:2012pu,daRocha:2017cxu}.
\par 
In this paper, we will scrutinize the Hawking flux of fermions across the event horizon of black holes that are solutions of static, spherically symmetric, asymptotically flat solutions in higher-derivative gravity with quadratic curvature terms, including quantum effects on the fermion dynamics predicted by the generalized uncertainty principle (GUP). \bltx{Some seminal works developed relevant aspects of GUP. Ref. \cite{Tawfik:2014zca} studied the GUP in the context of string theory, black hole physics and doubly special relativity, whereas bounds on the GUP parameter, based on  PLANCK observations on the cosmic inflation were discussed in Ref. \cite{Tawfik:2015rva}. Besides, Ref. \cite{Tawfik:2013uza} already predicted that GUP effects can drive black hole remnants, whose Hawking temperature, Bekenstein entropy, specific heat, emission rate and decay time were also calculated. GUP corrections to the entropy and thermodynamical quantities of charged black hole were derived in Ref. \cite{Tawfik:2015kga}, while GUP effects on compact stars were discussed in Ref. \cite{Ali:2013ii}. Besides, an interesting study of GUP and Lorentz violation was introduced in \cite{Tawfik:2012hz}. Other studies regarding the GUP and applications were scrutinized in Refs. \cite{gup51,gup1,gup22,gup3,gup4,SC}}. 

One may argue that if the evaporation process has an end, it will give rise to a remnant black hole. 
\bltx{In fact, an $s$-wave particle follows a trajectory outwards the black hole, that is classically forbidden. As a consequence of energy conservation, the black hole radius lessens as a function of the energy of the outgoing particle. This also provides a dynamical model of black hole radiation, since the mass of the black hole decreases, along the emission process outwards the event horizon \cite{k1}.}
Similarly, if the Hawking flux is extinguished, leaving a  black hole with vanishing quantum luminosity, then a remnant black hole may be produced. 
A black hole remnant consists of a black hole phase that evaporates under the Hawking radiation, which is either (absolutely) stable or long lived. The latter is also known as a metastable remnant \cite{Chen:2014jwq}.
The central concept involving black hole remnants consists of black holes whose size decreases during the Hawking evaporation process, reaching a minimal length, possibly near the Planck scale $l_p$ at which point the black hole ceases to evaporate. 
Notwithstanding the fact that  GUP effects in black hole remnants were already extensively studied in the literature \cite{Mu:2015qta,an1,Maluf:2018lyu,Chen:2013tha,Adler:2001vs,Chen:2014jwq}, a detailed analysis involving GUP in higher-derivative gravity is lacking, \bltx{despite a recent development \cite{Konoplya:2019ppy}}. We show, in particular, that the absolutely stable remnant case is attained for black hole solutions of quadratic gravity. We will compute higher-derivative corrections to the Hawking flux using the tunnelling method in a GUP context, governed by the GUP parameter $\upbeta$. 
\par
The paper is organized as follows:
in Sect. \ref{sss}, we will briefly review and discuss the black hole metric solution arising in higher-derivative 
gravity. Using the semiclassical approach of the WKB approximation, the tunnelling rate and the black hole luminosity 
 will be calculated in Sect.~\ref{SIII}, \bltx{showing that for appropriate parameters of the black hole solution  in higher-derivative 
gravity the black hole luminosity equals zero. It yields mini-black hole remnants}. Sect. is then devoted to the concluding remarks \ref{SIV}.
\section{Static spherically symmetric solutions in higher-derivative gravity}
\label{sss}
Motivated by the divergence structure appearing in the quantization of general relativity at one-loop, Stelle came up with a gravitational theory containing quadratic curvature invariants which turned out to be renormalizable \cite{Stelle:1976gc}, but saddly suffers from a ghost in its spectrum. Several solutions to the ghost issue have been proposed \cite{Lee:1969fy,Cutkosky:1969fq,Tomboulis:1977jk,Tomboulis:1983sw}, but no consensus has been reached so far. The ghost seems to be harmless at energies below the Planck scale \cite{Salles:2014rua}, which is the regime we are mostly interested in this paper. In any case, one can always project the ghost out by a suitable choice of boundary conditions \cite{Barnaby:2007ve,Kuntz:2019gup}.

The action of quadratic gravity 
\be
\label{acao}S=\int \mathrm{d}^4x\sqrt{-g}\left[\frac{1}{16\pi G} R - 2a R_{\mu\nu}R^{\mu\nu} + \left(b+{2a\over3}\right)R^2\right]\,,
\ee 
where $G$ denotes the 4D Newton constant, is both renormalizable and asymptotically free. In fact, the coefficients of the quadratic curvature terms vanish asymptotically in the ultraviolet regime of the theory. 
The action \eqref{acao} yields the EOMs \cite{Lu:2015psa}
\beq\label{movem}
B_{\mu\nu}  &=& T_{\mu\nu}\,,  \label{EOM11}
\eeq
where
\beq\label{tb}
B_{\mu\nu}& = & 
\frac{1}{8\pi G} G_{\mu\nu}-a \Box R_{\mu\nu}+\frac{1}{3}\left( a-3b \right) \mathit\nabla_{\mu} \nabla_{\nu} R
-2a R^{\rho\sigma}R_{\mu\rho\nu\sigma}  \notag\\
& &
+\frac14 g_{\mu\nu} \left[2 a R^{\rho\sigma}R_{\rho\sigma} +\frac{2}{3}\left( a +6b \right) \Box R-\left(\frac{2a}{3}+b \right) R^{2}\right]+\left( \frac{2a}{3}+b \right) R R_{\mu\nu},
\eeq
for $G_{\mu\nu}$ being the Einstein tensor.  The tensor (\ref{tb}), whose trace reads 
\begin{equation}\label{eq:EOMtrace}
B_{\;\mu}^{\mu} = -  \frac{1}{16\pi G}R +6 b \Box R  = T_{\;\mu}^{\mu}\,,
\end{equation}
satisfies the effective field equations  
\bltx{\beq
\nabla^{\nu}B_{\mu\nu} = 0.\label{bian}
\eeq }
When $b=0$, corresponding to the Einstein--Weyl theory, the sign of $a$ can be derived 
when one linearizes the  Minkowski metric, namely, $g_{\mu\nu}=\eta_{\mu\nu}+
h_{\mu\nu}$, yielding 
\begin{equation}
-\ft13a \Box \left(\Box - \frac{1}{16\pi G a}\right) h_{\mu\nu} =0\,.
\end{equation}
The range $a > 0$ implies a stable theory, in the sense that no tachyonic instabilities sets in. In addition, there are massive spin-2 and spin-0 excitations, respectively with masses $m_2^2 = \frac{1}{32\pi G a}$ and $m_0^2 = \frac{1}{96\pi G b}$. The former corresponds to the aforementioned ghost. Hence, one can write $B_{\;\mu}^{\mu} = 6 b \left(\Box  -m_0^2 \right)R$ \cite{Lu:2015psa}.  

Solutions of the EOM (\ref{EOM11}) were scrutinized in Ref. \cite{Lu:2015psa}, using 
the Frobenius procedure, with respect to the radial coordinate, $r$,  to implement indicial equations for the leading asymptotic behaviour as $r\rightarrow0$. Ref.\,\cite{Stelle:1977ry} derived the leading asymptotic profiles 
of the temporal and radial metric coefficients, 
\begin{eqnarray}\label{m21}
ds^2=-A(r)\, dt^{2}+B(r)\, dr^{2}+r^{2}d\theta^{2}+r^{2}\sin^{2}\theta\: d\phi^{2}\,,
\end{eqnarray} 
\bltx{In fact, one can express
\begin{equation}\label{888}
B_{\mu\nu}={\rm diag}(B_{tt}(r), B_{rr}(r),  B_{\theta\theta}(r), B_{\theta\theta}(r) \sin^2\theta),
\end{equation}
whose components are related by the radial component of (\ref{bian}):
\begin{equation}\label{bianr}
\left(\frac{B_{rr}}{A}\right)'+\frac{2B_{rr}}{Ar}+\frac{B'B_{rr}}{2AB}-\frac{2B_{\theta\theta}}{r^3}+\frac{B'B_{tt}}{2B^2} \equiv 0\,.
\end{equation}
For any given source, $T_{\mu\nu}$, the following equations govern the system,
\bse\label{212}
\begin{eqnarray}
B_{tt} & = &\, \frac12 T_{tt}\,, \label{bian1} \\
B_{rr} & = &\, \frac12 T_{rr}\,. \label{bian2}
\end{eqnarray}
\ese
Sourceless solutions, corresponding to $T_{\mu\nu}=0$, are aimed, hereon, 
\bse\label{bi}
\begin{eqnarray}
B_{tt} & = &\, 0\,, \label{bian3}\\
B_{rr} & = &\, 0 \,.\label{bian4}
\end{eqnarray}
\ese
Solutions of the coupled system (\ref{bi}) can be emulated as the asymptotic behaviour of solutions to the equations of motion was analysed near the origin \cite{Stelle:1977ry}. The temporal and radial metric components  (\ref{m21}) can be then expanded, in Frobenius series, as
\begin{equation}
\begin{split}\label{Frobseries}
A(r) & = a_nr^n+a_{n+1}r^{n+1}+a_{n+2}r^{n+2}+\ldots \,,\\
B(r) & = b_{m} \left( r^{m}+b_{m+1}r^{m+1}+b_{m+2}r^{m+2}+\ldots \right) \,,
\end{split}
\end{equation}
where $a_n, b_{m} \neq 0$. Replacing the series \eqref{Frobseries} into the EOMs \eqref{bi} and analysing the consistent possibilities for the $(m,n)=(1,-1)$ indexes yields the family of solutions \cite{Stelle:1977ry}. The corresponding equations can be, therefore, solved order by order, for the coefficients $a_{n},b_{n}$. 
Clearly, some of the coefficients are free parameters, not being possible to determine them. One example of 
a free parameter, in $B(r)$, is a scaling of the temporal coordinate.  The family of solutions read}
\begin{eqnarray}
A(r) &=&\zeta\left[\frac{1}{r}
+\alpha_1
+\alpha_2 r^2
+\frac{1}{16} r^3 \left(\alpha_1 \alpha_2+\alpha_1^4+\alpha_3\right)
-\frac{3}{40}  r^4 \left(\alpha_1 \left(\alpha_1 \alpha_2+\alpha_1^4+\alpha_3\right)\right)\right]\,,
\label{metrica12}\\
B(r) &=&\alpha_1 r
-\alpha_1^2 r^2
+\alpha_1^3 r^3
+\alpha_3 r^4
-\frac{1}{16} r^5 \left(\alpha_1 \left(3 \alpha_1 \alpha_2+19 \alpha_1^4+35 \alpha_3\right)\right).\label{metrica13}
\end{eqnarray}
\bltx{The parameter $\zeta=M/2$ in Eq. (\ref{metrica12}) plays the role of the Misner--Sharp mass. Refs. \cite{Lu:2015psa,Stelle:1977ry}
showed that the maximum number of integration-constant parameters that govern this family of  solutions comes from the fact that Eqs. (\ref{bian1}, \ref{bian2}) consist of a third-order coupled system of  non-linear ODE for the metric coefficients, having four free parameters, $\alpha_1, \alpha_2, \alpha_3, \zeta$.} As it is an expansion around $r=0$, the metric (\ref{metrica12}, \ref{metrica13}) is trustworthy for the computation of the Hawking radiation spectrum, as terms in order beyond $\mathcal{O}(r^6)$ are totally negligible.
The family (\ref{metrica12}, \ref{metrica13}) includes the standard Schwarzschild solution \cite{Stelle:1977ry}, as a solution of the higher-derivative EOMs. At the origin, the family of solutions (\ref{metrica12}, \ref{metrica13}) presents a physical singularity, as  $\lim_{r\to0}R_{\mu\nu\rho\sigma}R^{\mu\nu\rho\sigma}\sim r^{-6}$ \cite{Lu:2015psa,Stelle:1977ry}.
 This family also includes non-Schwarzschild black holes. 
\section{Hawking radiation spectrum, flux and black hole Hawking luminosity}
\label{SIII}
The GUP asserts that  $ 
\Updelta x\, \Updelta p
\gtrsim
\frac{\hbar}{2}\left[1+ \upbeta\, \Updelta p^2\right],$ 
for $\upbeta =\upbeta_0/\mpl^2$, being $\upbeta_0 $ a dimensionless parameter that accounts for  
effects of quantum gravity, having the bound $|\upbeta_0| \lesssim 10^{21}$  \cite{Das:2008kaa,Scardigli:2014qka}. In the GUP apparatus, $x_j = X_{j}$
and $p_j = P_j\,(1 + \upbeta\, p^2)$ are respectively position and momentum operators, where  $\left[X_{j},P_{k}\right]= i\, \hbar\, \delta_{jk}$. \bltx{Given the spacetime metric $g_{ij}$,} it implies that 
\be
g_{ij}p^i p^j
=
\hbar^2\,g^{jk}\partial_j \partial_k
\left(2\,\upbeta\, \hbar^2\, g^{pq}\,\partial_p\partial_q-1\right)
\ .
\label{eq2.3}
\ee
\bltx{Given the set $\{\upgamma^\mu\}$ of gamma matrices in spacetime, satisfying the Clifford--Dirac relation $\gamma^\mu\gamma^\nu+\gamma^\nu\gamma^\mu=2g_{\mu\nu}$,} the Dirac equation, governing fermions, with electromagnetic field $A^\mu$, reads 
\be
\left\{
i\,\upgamma^{\mu}
\left[\hbar\left(\nabla_{\mu}+\Upomega_{\mu}\right)
+i\,e\,\mathcal{A}_{\mu}\right]
+m\mathbb{I}_{4\times 4}
\right\}
\psi(x^\mu)
=0
\ ,
\label{eq2.6}
\ee
where $\Upomega _\mu=-\frac{i}{2}\,\upomega _\mu^{\,\rho\sigma}\,\Upsigma_{\rho\sigma}$, and
$\upomega _{\mu\; \sigma}^{\;\rho}=e_\nu^{\,\rho} e^\alpha_{\,\sigma} \upgamma^\nu_{\mu\alpha}$
is the spin connection. \bltx{Besides, $e$ denotes the electric charge in Eq. (\ref{eq2.6}), whereas $\mathcal{A}_{\mu}$ is the electromagnetic 
gauge potencial.}
Eq.~(\ref{eq2.3}) can be substituted into Eq.~(\ref{eq2.6}), together with 
 the energy of a particle of mass $m$ and electric charge $e$ on the mass shell,
$i\, \hbar\,\partial_0\left[
1 + \upbeta\left(p^2 + m^2\right)\right]$ \cite{Chen:2013tha}.  The Dirac equation then becomes 
\be
\left\{i\,\hbar\,\upgamma^{0}\partial_{0}
+
\left[m\mathbb{I}_{4\times 4}+\upgamma^{\mu}\left(
i\,\hbar
\Upomega_{\mu}+i\hbar^2\upbeta\partial_\mu
-e\mathcal{A}_{\mu}\right)
\right]
\left(
1+\upbeta(\hbar^{2}\,g_{pq}\,\partial^{p}\,\partial^{q}-m^{2})
\right)
\right\}
\psi(x^\mu)
=
0
\ .
\label{eq2.9}
\ee
\par
The black hole Hawking radiation for  fermions can be computed with the aid of the tunnelling procedure, where the fermion is assumed to have the following form, without loss of generality
~\cite{hoff}: 
\be
\Psi
=
\left(
\psi_1,0,\psi_2,0\right)^\intercal\,
e^{\frac{i}{\hbar}\,J(t,r,\theta,\phi)}
\ ,
\label{304}
\ee
for an action $J$ and wavefunctions ${\psi_1}$ and ${{\psi_2} }$. 
\bltx{It is worth emphasizing that employing the WKB approximation, the tunnelling probability for a classically forbidden trajectory of the $s$-wave outwards the horizon reads $\Upgamma\propto \exp(2\Im J)$, where $J$ is the classical action of the trajectory to leading order in $\hbar$. When one expands the action in terms of the particle energy, the Hawking temperature is recovered at linear order. In fact,  for $2I = \beta E + \mathcal{O}(E^2)$, it yields $\Upgamma\propto \exp(2\Im J)\approxeq \exp(\beta E)$. This corresponds to the regular Boltzmann factor  for emission at the Hawking temperature, for a particle of energy $E$, for $\beta=1/T$, where $T$ is black hole horizon temperature.  Higher order terms regard self-interaction \cite{Kraus:1994by}. To compute the black hole temperature linear order expansion suffices. The tunnelling method 
makes it possible to calculate the imaginary part of the action for the emitted particle. The protocol to be prescribed consists of assuming a Hamilton--Jacobi-like ansatz that regards spin-1/2 fermions. From the symmetries of the metric that describes the black hole geometry, the form of the action, $J$, will be chosen by an appropriate ansatz. This procedure is based on applying the WKB approximation to the Dirac equation \eqref{eq2.9}, corrected by GUP effects, that governs spin-1/2 fermions.}
The metric~(\ref{m21}) yields the tetrads
\be\label{tetr}
e_\mu^{\,\alpha}
=
{\rm{diag}}
\left[
\sqrt{A(r)}, \sqrt{B(r)}, r, r\,\sin\theta
\right].
\ee
Since the method for computing the tunnelling rate is representation-independent, the one used hereon is more appropriate, for $\upgamma^5=i\upgamma^0\upgamma^1\upgamma^2\upgamma^3$, where the $\upsigma_i$ denote the Pauli matrices: 
\ba
&&
\upgamma_{t}
=-i\sqrt{A(r)}\upgamma^5,
\quad
\upgamma^{\theta}
=
r
\begin{pmatrix}
\mathcal{O}&\upsigma_1\\
\upsigma_1&\mathcal{O}
\end{pmatrix},\quad
\upgamma^{r}
=
\sqrt{B(r)}
\begin{pmatrix}
\mathcal{O}&\upsigma_3\\
\upsigma_3&\mathcal{O}
\end{pmatrix},
\quad
\upgamma^{\phi}
=
r\sin\theta
\begin{pmatrix}
\mathcal{O}&\upsigma_2\\
\upsigma_2&\mathcal{O}
\end{pmatrix}.
\label{305}
\ea
Eqs.~(\ref{304}, \ref{305}) replaced in the GUP-corrected Dirac equation, (\ref{eq2.9}), yields the following EOMs, using the WKB regime to order in $\hbar$:
\ba
&&
{\psi_1}
\left\{
\frac{i}{\sqrt{A}}
\left[
\dot{J}\!-\!e\,\mathcal{A}_t\left(1\!-\!\upbeta (m^{2}+\kappa)\right)
\right]
-m
\left(1-\upbeta(m^{2}-\kappa)\right)
\right\}
=
{\psi_2}
\left(
1-\upbeta(m^{2}-\kappa)
\right)
\frac{J'}{\sqrt{B}},
\quad
\label{306}
\\
&&
{\psi_2}
\left\{
\frac{i}{\sqrt{A}}
\left[\dot{J}
+e\,\mathcal{A}_t
\left(1-\upbeta (m^{2}+ \kappa)\right)
\right]
+m
\left(1-\upbeta (m^{2}-\kappa)\right)
\right\}
=
-\psi_1\left(1-\upbeta(m^{2}-\kappa)\right)
\frac{J'}{\sqrt{B}},
\quad
\label{307}
\\
&&\,\;\quad\quad\qquad\qquad\qquad\qquad\qquad\left(1-\upbeta(m^2+\kappa)\right)
\left(
\frac{\partial }{\partial\theta}
+\frac{i}{\sin\theta}\frac{\partial }{\partial  {\phi}}
\right)J\label{388}
=
0
\ ,
\ea
with $
J'=\frac{\partial J}{\partial r}$, $\dot{J}=\frac{\partial J}{\partial t}$, and
\be
\kappa
=
B\,J'^2
+
\frac{1}{r^2}\left(\frac{\partial J}{\partial\theta}\right)^2
+
\frac{1}{r^2\sin^2\theta}\left(\frac{\partial J}{\partial\phi}\right)^2
\,.
\ee 
Expressing the action as  
\be
J
=
-\upomega\, t
+w(r)
+{\mathit{\Theta}}(\phi, \theta)
\ ,
\label{3010}
\ee
with $\upomega$ denoting the energy of the emitted fermionic spectrum, the tunnelling rate will be then computed ~\cite{2,Vanzo:2011wq,an1}.
Substituting Eq.~(\ref{3010}) in  
Eq.~(\ref{388}) yields 
\be
\left(\frac{\partial \mathit{\Theta}}{\partial\theta}
+\frac{i}{\sin\theta}\frac{\partial \mathit{\Theta}}{\partial\phi}\right)
\left[
\upbeta \left(B\,(w')^2
+
 \frac{1}{r^2}\left(\frac{\partial \mathit{\Theta}}{\partial\theta}\right)^2
+
 \frac{1}{r^2\,\sin^2\theta}\left(\frac{\partial \mathit{\Theta}}{\partial\phi}\right)^2
+ m^2\right)
- 1\right] =0.
\label{3011}
\ee
As the part of the equation, that is in the inner side of the square brackets, will be not identically null, one must have
\be
\left(\frac{\partial}{\partial\theta}
+\frac{i}{\sin\theta}\frac{\partial}{\partial\phi}\right)\mathit{\Theta}
=
0.
\label{eqTheta}
\ee
This means that  the ${\mathit{\Theta}}$ function will not contribute for the tunnelling process.
Now, replacing Eqs.~(\ref{3010}, \ref{eqTheta}) into Eqs.~(\ref{306}, \ref{307}),
yields  
\be
\xi_0+\xi_1w'^{2}+\xi_2w'^{4}+\xi_3w'^{6}
=
0
\ ,
\label{3014}
\ee
where
\begin{subequations}
\ba
\xi_0
&=&
-\left[e^2 \mathcal{A}_t^2+m^{2}\,A\right]
\left(1-\upbeta\, m^{2}\right)^2
+2\,\upomega\, e\,\mathcal{A}_t
\left(1-\upbeta\, m^{2}\right)-\upomega^2
\ ,
\\
\xi_1
&=&
\frac{\upbeta}{ B}
\left\{A\left(1-\upbeta^2\,m^4\right)
+2\, e\,\mathcal{A}_t
\left[e\,\mathcal{A}_t
\left(1-\upbeta\, m^2\right)
-\upomega\right]
\right\}
\ ,
\\
\xi_2
&=&
A
\left(2-\upbeta\,m^{2}\right)-\frac{\upbeta^2}{ B^2}
 e^2 \mathcal{A}_t^2
\ ,
\\
\xi_3
&=&
\upbeta^{2}\frac{A}{B^{3}}\,
\ .
\label{3015}
\ea
\end{subequations}
In what follows the scaling $\zeta\mapsto\frac{\zeta}{\mpl}$, $\alpha_1\mapsto \alpha_1\mpl$, $\alpha_2\mapsto \alpha_2\mpl^3$ and $\alpha_4\mapsto \alpha_4\mpl^4$ is more illustrative and will be adopted. 

\bltx{Let us denote by $w_\pm$ the corresponding motion away from ($+$) and toward ($-$) the black  hole horizon.
The ${}_\pm$ cases correspond to
outgoing/incoming solutions of the same spin. Note that neither of these cases
is an antiparticle solution since we assumed positive frequency modes as a part
of the ansatz.  In computing the imaginary part of the action, both $\Theta$ and $w_\pm$ are, in general, complex functions.
Therefore, they will contribute for the emission probability, $\Upgamma$, defined to be the probability of outward scattering-to-probability of inward scattering ratio.}

Solving
Eq.~(\ref{3014}) on the event horizon yields the imaginary part of the action, 
\be\label{imm}
\Im\, w_\pm (r)
=
\pm \frac{\pi}{4}\,
\frac{r_+^2\,\upomega
\left(1+ \upbeta\,\Uplambda\right)}
{r_+ - \chi\,r_-}
\ ,
\ee
where
\begin{subequations}
\begin{eqnarray}
\chi
&=&
\frac{2\zeta\left(3\, \zeta^3/\mpl^3+\zeta/\mpl\right)}{\mpl^5\left(5\zeta\alpha_1^2\alpha_2+\alpha_3\right)\left(\frac{\zeta}{2\mpl}+g_{\alpha}\right)}
\ ,
\label{a}
\\
\Uplambda
&=&
\frac{3}{2}\,m^2
+{\frac{e\,m^2\,\mathcal{A}_t}{\upomega-e\,\mathcal{A}_t}}
-\frac{12\,e^2\,\mathcal{A}_t\mpl^6\left({2}(\zeta^4\alpha_1^3 \alpha_2\alpha_3)+{3}\alpha_1^5
\alpha_2^2\right)}{{\zeta\mpl^{11}}+{8}{\zeta^2\alpha_1^4 \alpha_2^2\alpha_3}+{9\mpl^3}{\zeta\alpha_1^3 \alpha_2\alpha_3}}
+\frac{\zeta\upomega}{\zeta/\mpl+g_\alpha}\,,
\label{3017}
\end{eqnarray}
\end{subequations}
for
\beq
g_\alpha&=& \frac{2}{\mpl^{16}}{\zeta^4\alpha_1^6 \alpha_2^2\alpha_3^2}+\frac{4}{\mpl^{8}}{\zeta^2\alpha_1^3 \alpha_2\alpha_3}\nonumber\\
 &&+\left[\left(\alpha_1^2\mpl^2 r \left(15 \alpha_2\mpl^3 r^3+32\right)+95 \alpha_1^5\mpl^5 r^4+\alpha_1\mpl \left(175 \alpha_3\mpl^4 r^4-16\right)-64 \alpha_3\mpl^4 r^3\right.\right.\nonumber\\&&\left. \left.-48 \alpha_1^3\mpl^3
   r^2\right)\left(r^3 \left(3 r \left(\alpha_1 \alpha_2+\alpha_1^4+\alpha_3\right)\mpl^4 \left(8 \alpha_1\mpl r-5\right)-160 \alpha_2\mpl^3\right)+80\right)\right]^{1/2}\big\vert_{r=r_+}\label{gal}
\eeq
and 
\beq\label{jh}
r_\pm&=& -\frac{4 \alpha_2}{\alpha_3\mpl}\pm\sqrt{2} \sqrt{\frac{8 \alpha_2^2}{\alpha_3^2\mpl^2}+\frac{f_{\alpha_2,\alpha_3}}{3^{2/3}
   \alpha_3\mpl^4}+\frac{1}{\sqrt[3]{3} f_{\alpha_2,\alpha_3}}}\nonumber\\&&+\frac{1}{2} \left(\frac{256 \alpha_2^3}{\alpha_3^3 \sqrt{\frac{4
   \alpha_2^2}{\alpha_3^2\mpl^2}+\frac{\sqrt[3]{3} \alpha_3\mpl^4+f_{\alpha_2,\alpha_3}^2}{6 \alpha_3\mpl^4 \sqrt[3]{12 \alpha_2^2\mpl^6+\sqrt{144
   \alpha_2^4\mpl^{12}-\frac{\alpha_3^3}{3}}}}}}-\frac{128 \alpha_2^2}{\alpha_3^2\mpl^2}+\frac{8 f_{\alpha_2,\alpha_3}}{3^{2/3} \alpha_3}+\frac{8}{\sqrt[3]{3}f_{\alpha_2,\alpha_3}}\right).
   \eeq Eq. (\ref{jh}) is displayed for $\alpha_1=0$,
where $f_{\alpha_2, \alpha_3}=\sqrt[3]{36 \alpha_2^2\mpl^6+\sqrt{1296 \alpha_2^4\mpl^8-3 \alpha_3^3\mpl^{12}}}$. 
For $\alpha_1\neq 0$, the general solution, having dozens of pages, is opted not to be displayed here.

Thus, the tunnelling rate of fermions reads 
\be
\Upgamma
\simeq
\frac{e^{- 2\,\Im {{\mathit{\Theta}}}-2\,\Im w_+}}
{e^{-2\,\Im {{\mathit{\Theta}}}-2\,\Im w_-}}
\simeq
\exp\!\left[-\frac{8\,\pi\,M\left(1 + \upbeta\,\Uplambda\right)\upomega}
{\mpl^2 }
\right].
\label{3018}
\ee
As $\zeta=M/2$ in Eq. (\ref{metrica12}), in particular for Schwarzschild-like black holes, the tunnelling rate (\ref{3018}) has the proper form $\Upgamma=\exp\!\left(-\frac{8\,\pi\,M\upomega}
{\mpl^2 }\right)$ when no GUP effects are considered, i.e., when $\upbeta\to0$. 

In what follows, for the sake of simplicity, one takes $\mathcal{A}_t=0$ and express 
\be
\upbeta\,\Uplambda
=
\upbeta_0
\left(
\frac{3\,m^2}{2\,\mpl^2}
+\frac{\zeta\upomega}{\zeta/\mpl+g_\alpha}
\right).
\ee
The tunnelling rate of evaporation, $\Upgamma$, is plotted in Figs.~\ref{b1}-\ref{bs3} for various cases. 
\par
Taking into account that $\upomega\sim\mpl^2/\zeta$, the tunnelling rate of evaporation~\eqref{3018} can be written as the Boltzmann term 
$\Upgamma=\exp\left(-{\upomega}/{T}\right)$, where
\be
T
=
\frac{\mpl^2}{4\,\pi\,\zeta\left(1+\upbeta\,\Uplambda\right)}
\ .
\label{3019}
\ee
It is worth emphasizing  that \begin{eqnarray}
T_0
&=&
\frac{\hbar}{4\,\pi} \left[\sqrt{A'(r)(B^{-1})'(r)}\right]_{r=r_\pm}\nonumber\\
&=&\frac{\mpl^2}{64\sqrt{5}\pi \zeta}+\left[\left(\alpha_1^2\mpl^2 r \left(15 \alpha_2\mpl^3 r^3+32\right)+95 \alpha_1^5\mpl^5 r^4+\alpha_1\mpl \left(175 \alpha_3\mpl^4 r^4-16\right)-64 \alpha_3\mpl^4 r^3\right.\right.\nonumber\\&&\left.\left.-48 \alpha_1^3\mpl^3
   r^2\right)\left(r^3 \left(3 r \left(\alpha_1 \alpha_2+\alpha_1^4+\alpha_3\right)\mpl^4 \left(8 \alpha_1\mpl r-5\right)-160 \alpha_2\mpl^3\right)+80\right)\right]^{1/2}\big\vert_{r=r_+}
\label{T0}
\end{eqnarray}
is the Hawking temperature of the black hole (\ref{metrica12}, \ref{metrica13}), obtained with the tunnelling method~\cite{k0}. 
The tunnelling rate~\eqref{3018}, thus, coincides to the Hawking standard one for black holes
with a sufficiently large mass, $M=\zeta/2$, such that the GUP correction is insignificant. 
\par
When the black hole mass is near the Planck scale, the $\Uplambda$ function (\ref{3017}) 
depends on $\upomega\sim M=\zeta/2 \sim \mpl$. In this regime, the fermion mass is clearly negligible. 
\bltx{For particles emitted in a wave mode labelled by energy $\omega$ and $\ell$, the probability for a black hole 
to emit a particle is equal to $\exp\left( -\frac{\upomega}{T}\right)$ times the probability for a black
hole to absorb a particle in the same mode, 
where $T$ is the black hole temperature. Balance condition demands that the ratio of the
probability of having $N$ particles in a particular mode 
to the probability of having $N-1$ particles in the same mode is $\exp\left(
-\frac{\omega}{T}\right).$ Hence, the average number $n_\ell(\upomega)$ in the mode %
$
n_\ell(\upomega)=n\left(  \frac{\upomega}{T}\right)$ can be derived, 
where $n\left(x\right)  =\frac{1}{\exp x+1},$ for fermions. 
Ref. \cite{Page:1976df} counted the number of modes, per frequency range, 
with periodic boundary conditions, around a black hole. The expected number emitted per mode $n_\ell(\upomega)$, to the average emission rate per frequency range, reads  
\begin{equation}
\frac{d n_\ell(\upomega)}{dt}=n_\ell(\upomega)\frac{\partial\omega}{\partial p_{r}%
}\frac{dp_{r}}{2\pi\hbar}=n_\ell(\upomega)\frac{d\omega}{2\pi\hbar},
\label{1238}%
\end{equation}
where $\frac{\partial\omega}{\partial p_{r}}$ is the radial velocity of the
particle, whereas the number of modes in the range $\left(
p_{r},p_{r}+dp_{r}\right)  $ is given by $\frac{dp_{r}}{2\pi\hbar}$, where
$p_{r}=\frac{\partial J}{\partial r}$ is the radial wavevector. 
}

Besides,  also the temperature is dependent on $\upomega\sim \mpl$. To carry out this dependence, let us consider Hawking fermions of energy $\upomega$ in some given mode $\ell$. Their emission probability can be described by the rate
${\Upgamma(\upomega)}=e^{-{\upomega}/{T(\upomega)}}$, up to a factor that encodes the absorption probability of the fermions by the black hole.
To quantify this reasoning, one denotes the average number of fermions carried by each $\ell$ mode, 
$
n_{\ell}(\upomega)
=
({1+\exp(\upomega/T)})^{-1}
=
\frac{\Upgamma(\upomega)}{1+\Upgamma(\upomega)}.$  
\bltx{Since each particle carries
off the energy $\omega$, the total luminosity is obtained from $\frac
{dn_\ell(\upomega)}{dt}$ by multiplying by the energy $\omega$ and summing up over
all energy $\omega$ and $\ell$,
\begin{equation}\label{lumi}
L=%
\frac{1}{2\pi\hbar}{\displaystyle\sum\limits_{\ell=0}^\infty}
\left(  2\ell+1\right)  \int^\infty_{0}\upomega n_\ell(\upomega){d\upomega}.
\end{equation}
 However, some of the radiation emitted by the horizon
might not be able to reach the asymptotic region. One needs to consider the
greybody factor $\left\vert G_{\ell}\left(  \omega\right)  \right\vert ^{2}$,
where $G_{\ell}\left(  \omega\right)  $ represents the transmission coefficient
of the black hole barrier which in general can depend on the energy $\omega$
and angular momentum $l$ of the particle. Therefore, black hole luminosity reads } 
\be
L
=\frac{1}{2\pi\hbar}
{\displaystyle\sum\limits_{\ell=0}^\infty}
\left(  2\ell+1\right)
\int^\infty_{0}
\upomega\,n_\ell(\upomega)\,\left\vert G_{\ell}\left(  \upomega\right)  \right\vert^{2}\d\upomega
\ , 
\ee
where $G_{\ell}\left(  \upomega\right)$ denotes the grey-body factors.  
For small black holes, when $\mpl^2/M\ll\upomega$, in the continuum limit,  the luminosity reads
\be
L
=
\frac{T^{4}\,\zeta^{2}}{2\pi\mpl\,l_p}
\int_0^{\infty}\left(\frac{\upomega}{T(\upomega)}\right)^{3}d\!\left(\frac{\upomega}{T(\upomega)}\right)
\int_0
n
\left[
\frac{\upomega}{T(\upomega)}
\left(  1+\frac{\ell\left(\ell+1\right)\mpl^{6}}{16\,\pi^2\, \upomega\,T(\upomega)\, {\zeta^4}}\right)
\right]
d\!\left(\frac{\ell(\ell+1)}{\zeta^2\,\upomega^2}\right).
\label{lum222}
\ee
Modelling the black hole by a sphere introduces an upper bound on the absorbed $\ell$ modes, given by $\ell(\ell + 1)\,\mpl^4\lesssim \frac{27}{4}\zeta^2\,\upomega^2$~\cite{Mu:2015qta}, as the $\ell$ modes beyond this range will not constitute the absorption spectrum of the black hole.
\par
{\color{black}
\par
\bltx{Applying the metric (\ref{m21}) with coefficients (\ref{metrica12}, \ref{metrica13})}, for $M\simeq\mpl$, the flux is given by
\be
\label{flux}
L(\zeta,\upbeta_0,m)
\approxeq
\frac{\pi \,\mpl^3}{2\,l_p\,\zeta^2}\left[1+\frac{\alpha_1^2(\alpha_1\alpha_2-\alpha_3)}{\alpha_1^2\alpha_3-\alpha_2^2}\right]
+\upbeta_0\,\frac{\mpl}{l_p}
\left(\frac{\pi \,\mpl^2}{2\,\zeta^2 g_\alpha}-\frac{\zeta-6\,\mpl}{18 \left(\alpha _1 \alpha _2+\alpha _3\right)8\,\mpl}F_{\alpha}\right)
\ ,
\ee
where 
\ba
F_\alpha
&=&
{2 \sqrt[3]{2} \alpha _1^6\mpl^7}{\sqrt[6]{{\alpha _2 \left(\alpha _1 \alpha _2+\alpha _3\right){}^2 \mpl^9\left(4 \alpha
   _1^9+243\alpha _2(\alpha _2^2 \alpha _1^2+2\alpha _2 \alpha _3 \alpha _1+\alpha _3^2)\right)}}}\nonumber\\
 &&  -243\alpha _2\mpl^{11}( \alpha _2^2 \alpha _1^2+2\alpha _2\alpha _3 \alpha _1- \alpha _3^2)-2 \alpha _1^9\mpl^9-2 \alpha _1^3\mpl^3\nonumber\\&&
 +2^{2/3}\mpl^{21/2}\left[\sqrt{\alpha _2 \left(\alpha _1 \alpha _2+\alpha _3\right){}^2 \left(4 \alpha _1^9+243 \alpha _2(\alpha _2^2 \alpha _1^2+2 \alpha _2 \alpha _3 \alpha _1^2+ \alpha _3^2)\right)}\right.\nonumber\\&&\left.-2 \alpha _1^9\mpl^9-243\mpl^{8}(\alpha _2^3 \alpha _1^2+ \alpha _3^2)\right]^{1/3}\label{flux1}
\ea
The black hole evaporation rate $\dot M\simeq -L$, with regards to the radiation flux~\eqref{flux}, leads to the Hawking
temperature for $\upbeta_0=0$ and vanishes for a certain value $M=M_0$.

\bltx{Before proceeding to the determination of black hole remnants, it is important to realize that the  parameters, $\alpha_1, \alpha_2, \alpha_3,$ in the metric coefficients (\ref{metrica12}, \ref{metrica13}) are (a priori) free, as 
 integration-constant parameters that govern the family of  solutions of Eqs. (\ref{bian1}, \ref{bian2}). 
 However, to shed new light on physical aspects of these solutions, one can constrain the parameters $\alpha_1, \alpha_2, \alpha_3$. In fact, one can implement, for example, the classical tests of GR in the Solar system, further probing physical constraints on these free parameters. Hence, the perihelion precession of Mercury, the deflection of light by the Sun
and the radar echo delay observations, consisting of well known tests for several solutions in GR, can be here applied in the context of the metric coefficients (\ref{metrica12}, \ref{metrica13}), to observationally and experimentally constrain the free parameters. Considering the light speed $c=2.998\times 10^{8}~{\rm m/s}$, the Solar mass 
$M_{\odot}=1.989\times 10^{30}~{\rm kg}$, let one regards the
motion of a planet on a Keplerian ellipse with semi-axes 
$a_1$ and $a_2$, where $a_2 = a_1\sqrt{1-e^2}$, where $e=~0.205615$ is the eccentricity of the orbit; 
$a_1=57.91\times 10^{9}~{\rm m},$ the Sun radius is 
$R_{\odot}=6.955 \times 10^8~ {\rm m}$ and the Newton's gravitational constant reads $G=6.67\times 10^{-11}~{\rm m^{3}kg^{-1}s^{-2}}$. 
Therefore, the classical tests of GR \cite{Bohmer:2009yx,Casadio:2015jva}, applied to the metric (\ref{m21}) with metric coefficients (\ref{metrica12}, \ref{metrica13}), yield a lower bound for the parameters $\alpha_1, \alpha_2, \alpha_3$, given by
\beq
|\alpha_1\alpha_2|\lesssim |59.28\pm 73.04|\;{\rm m}^{-3},\label{bb1}
\eeq for the perihelion precession, whereas
\beq
|\alpha_1\alpha_3|\lesssim |17.37\pm 69.48|\;{\rm m}^{-4}, \label{bb2}
\eeq for the light deflection. Finally, the radar echo delay analysis yields the bound
\beq
|\alpha_2\alpha_3|\lesssim |37.02\pm 40.21| \;{\rm m}^{-4}.\label{bb3}
\eeq
Due to the form of the metric coefficients (\ref{metrica12}, \ref{metrica13}), it is not possible to obtain separate bounds for each one of the parameters $\alpha_i$.}

For $\upbeta_0>0$, the GUP term in Eq.~\eqref{flux} can attain negative values, thus compensating 
the Hawking radiation for some values of $\alpha_1$, $\alpha_2$ and $\alpha_3$. 
In particular, for  $0<\upbeta_0\ll 1$, the Hawking flux vanishes for
\be
\bltx{\!\!\!\!\!\!\!\!\!\!\!\!\!\!\!\!\!M_0(\upbeta_0)\approxeq
3\upbeta_0^{-1/2}\,\mpl,\quad\qquad\text{for\;\;\; $\alpha_1=-3.419,\;\;\; \alpha_2=1.673$, \;\;\;$\alpha_3=2.137.$}}
\label{critical1}
\ee
When $\upbeta_0\approxeq 1$, 
the Hawking flux is equal to zero when 
\be
\bltx{\!\!\!\!\!\!\!\!\!\!\!\!\!\!\!\!M_0(\upbeta_0)\approxeq
3.7\mpl,\label{critical12}
\quad\qquad\text{for\;\;\; $\alpha_1=2.316,\;\;\; \alpha_2=12.420$,\;\;\; $\alpha_3=-0.219.$}}
\ee 
For  $\upbeta_0\gg 1$, the critical mass, \bltx{for which the Hawking flux vanishes, producing  black hole remnants,} reads  
\be
\bltx{\!\!\!\!\!\!\!\!\!\!\!\!\!\!\!M_0
\sim
7.1\mpl,\quad\qquad\label{critical2}\text{for\;\;\; $\alpha_1=-5.372,\;\;\; \alpha_2=11.003, \;\;\;\alpha_3=2.914$.}}
\ee
 \bltx{These particular values of the $\alpha_i$  parameters, in Eqs. (\ref{critical1}) -- (\ref{critical2}), are in full agreement with the 
physical bounds (\ref{bb1}) -- (\ref{bb3}).}

It is worth noticing that for $\upbeta=0$ the Hawking flux does not vanish, whatever the values of $\alpha_1, \alpha_2, \alpha_3, \zeta$ are taken in the metric (\ref{metrica12}, \ref{metrica13}).

\begin{figure}[htb!]
\begin{minipage}{14pc}
\includegraphics[width=18pc]{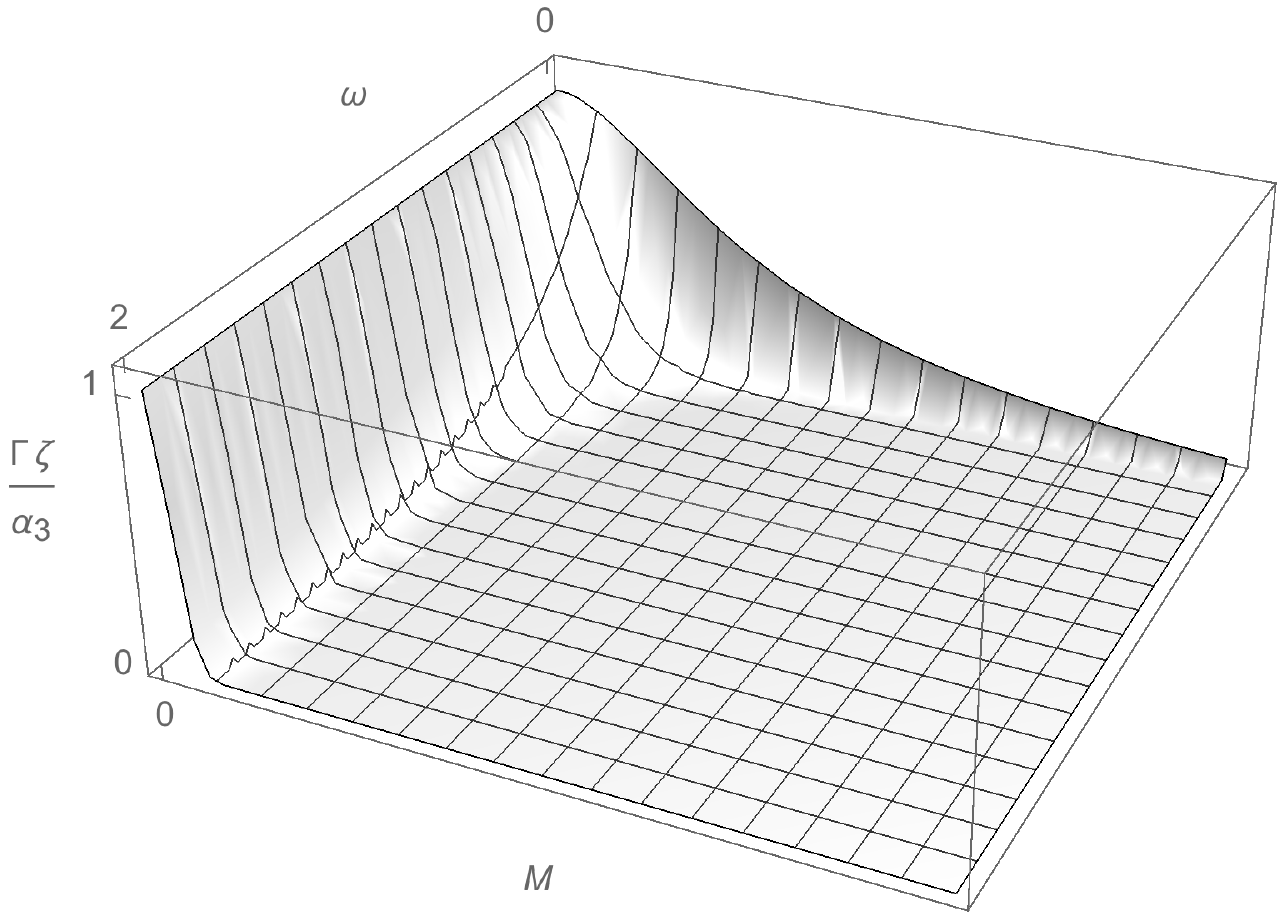}
\caption{\label{b1}\footnotesize{Hawking radiation spectrum $\Upgamma$ (normalized by the $\frac{\zeta}{\alpha_3}$ ratio) with respect to both $\upomega$ and the black hole mass $M=\zeta/2$ (in powers of $\mpl$), for $\upbeta_0=10^5$ and $\alpha_1=\alpha_2=1$. }}
\end{minipage}\hspace{5pc}
\begin{minipage}{14pc}
\includegraphics[width=17.5pc]{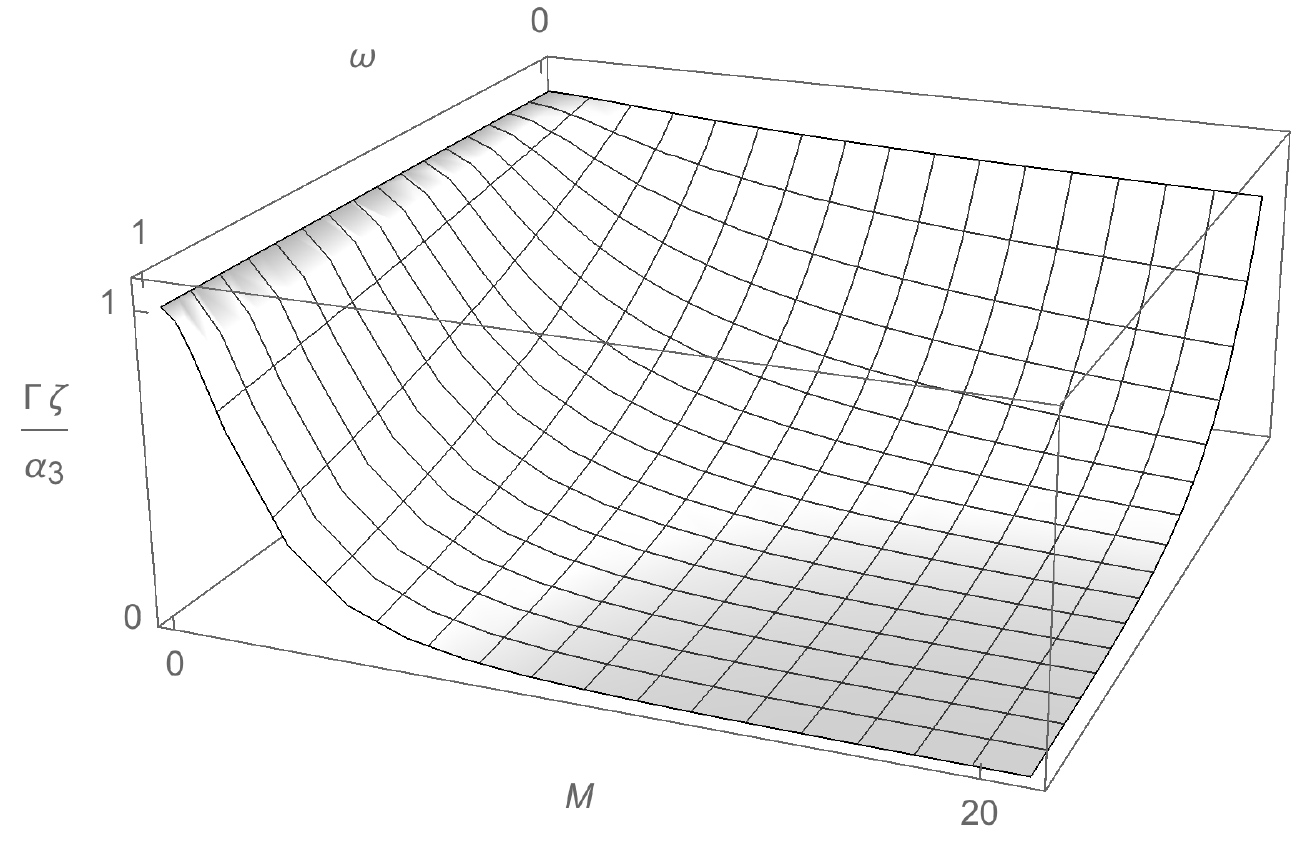}
\caption{\label{b2}\footnotesize{Hawking radiation spectrum $\Upgamma$ (normalized by the $\frac{\zeta}{\alpha_3}$ ratio), with respect to both $\upomega$ and the black hole mass $M=\zeta/2$ (in powers of $\mpl$),
for $\upbeta_0=0=\alpha_1=\alpha_2$.
}}
\end{minipage}
\begin{minipage}{14pc}
\includegraphics[width=16pc]{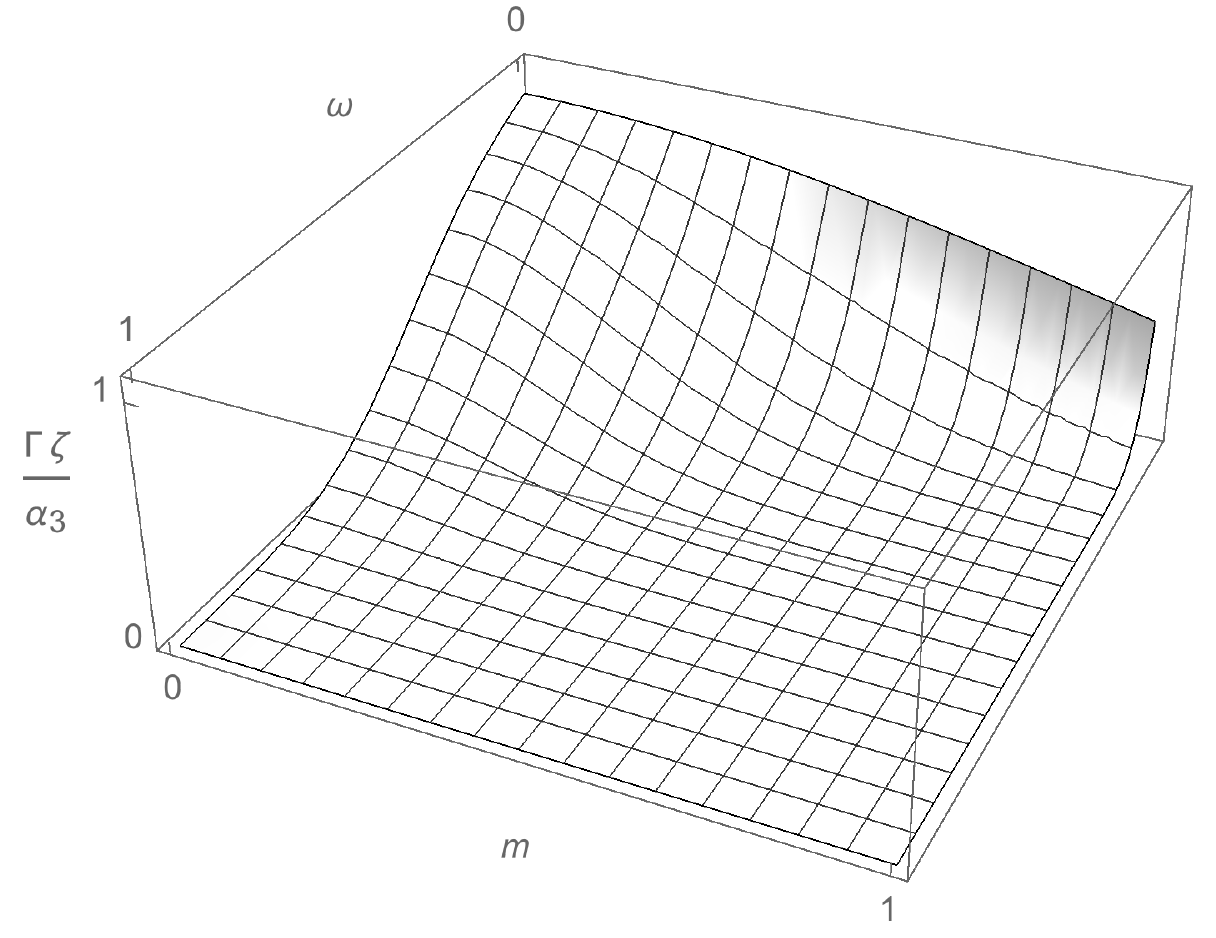}
\caption{\label{b3}\footnotesize{Boltzmann factor, with respect to both  $\upomega$ and the fermion mass $m$, for $\upbeta_0=10^5$ and $\alpha_1=\alpha_2=1$.}}
\end{minipage}\hspace{5pc}
\begin{minipage}{14pc}
\includegraphics[width=15.5pc]{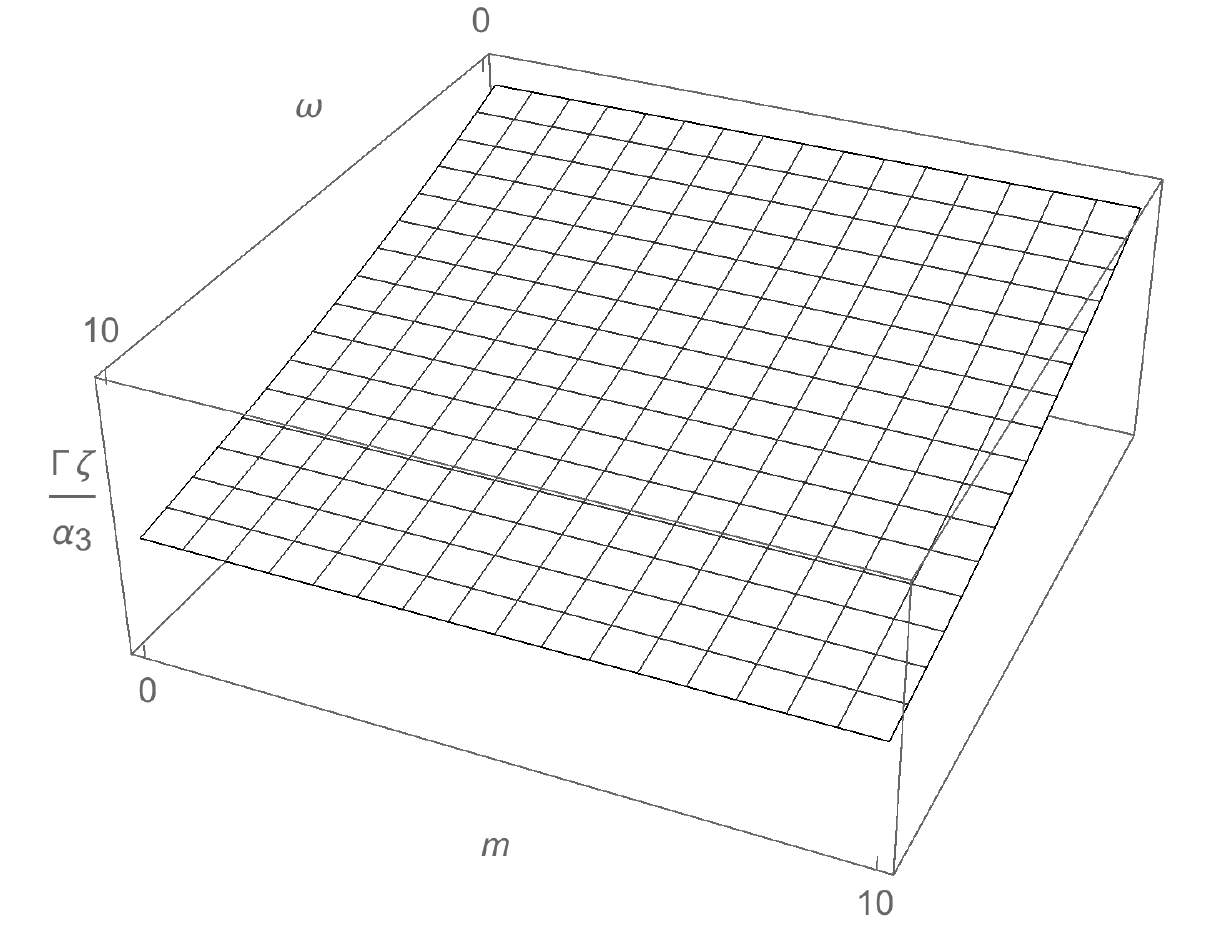}
\caption{\label{b4}\footnotesize{Hawking radiation spectrum $\Upgamma$ (normalized by the $\frac{\zeta}{\alpha_3}$ ratio), with respect to both $\upomega$ and the 
fermion mass $m$, for $\upbeta_0=0=\alpha_1=\alpha_2$.
}}
\end{minipage}
\includegraphics[width=17pc]{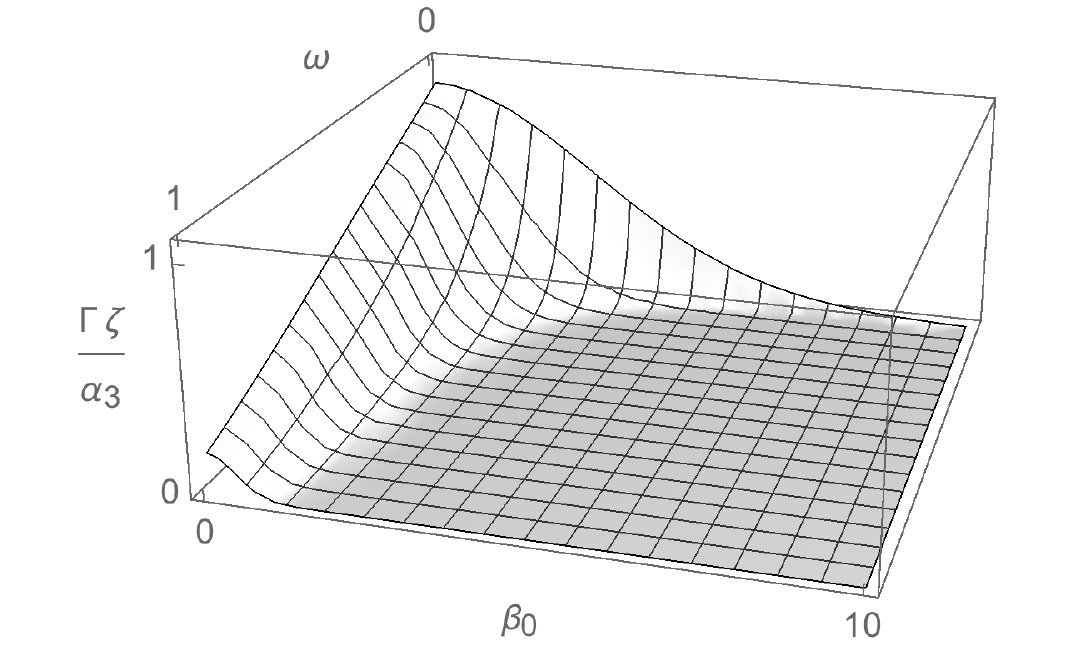}
\caption{\label{bs3}\footnotesize{Hawking radiation spectrum $\Upgamma$ (normalized by the $\frac{\zeta}{\alpha_3}$ ratio), with respect to both $\upomega$ and $\upbeta_0$, for $\alpha_1=\alpha_2=0.1$. 
}}
\end{figure}

\bltx{A crucial question when one tries to construct a quantum theory of gravity is its renormalizability. 
A conservative approach to obtain a renormalizable theory of gravity consists of adding higher-order terms to the theory. Among all possibilities of finding a renormalizable model, an action with curvature invariants up to second order stands out for its simplicity. Such theory is described by the action (\ref{acao}), which is equivalent to the theory studied in Ref. \cite{Stelle:1976gc}}
\beq
\label{acao2}
S = \int d^4x\sqrt{-g}\left(\frac{1}{16 \pi G} R + b R^{2}-
a C_{\mu\nu\rho\sigma}C^{\mu\nu\rho\sigma}\right)\,,
\eeq
for $C_{\mu\nu\rho\sigma}$ being the Weyl tensor components, due to the Gauss-Bonnet invariant.

As the Frobenius expansion was obtained near the origin in (\ref{metrica12}, \ref{metrica13}), it is now opportune to analyze the possibility of black hole remnants for an expansion around a nonzero radius $r_h$. Ref. \cite{Stelle:1976gc} showed that in static and asymptotically flat backgrounds, the existence of an event horizon yields a vanishing 
Ricci scalar. \bltx{Therefore the Schwarzschild solution is also the solution of the (\ref{acao2})}. Hence, the quadratic theory consists of Einstein--Weyl gravity, with action (\ref{acao2}) 
with $b=0$.  The corresponding EOMs read, for the metric \eqref{m21}, 
\beq
&&r^2 \left[ A \left(2  \frac{A''}{B}-\frac{A'B'}{B^2}\right)- A'^2\right]
+4   A \left[r f A'-A \left(\frac{B'}{B^2}r -\frac{1}{B}+1\right)\right]
\label{EOM1}=0
\\
&&a  \Bigg\{
r^3  \frac{A'}{B} \left[ -A \frac{A'B'}{B^2}+ \frac{A'^2}{B}+2 A^2\left(\frac{2B'^2}{B^3}-\frac{B''}{B^2}\right) \right]-\frac{A}{B^2} \left(3 r^2 A'^2+8 A^2\right)
\nonumber\\&&-2 \frac{A^2}{B}  \left(-r^2 A' \frac{B'}{B^2}-2 A \left(r^2 \left(\frac{2B'^2}{B^3}-\frac{B''}{B^2}\right)+r \frac{B'}{B^2}+2\right)\right)
-r A^3 \frac{B'}{B^2} \left(3 r \frac{B'}{B^2}+4\right)
\Bigg\}
\nonumber\\&&
+\frac{1}{8\pi G}  r^2 B \left(r{A'}+A (1-B)\right)=0\,.
\label{EOM2}
\eeq
The Schwarzschild standard metric, considering $A(r) = 1/B(r) = 1-\frac{2M}{r}$, for $r_h=2M$ in this case, 
satisfies the EOMs (\ref{EOM1}, \ref{EOM2}). 

The quadratic gravity theory does not present an analytical solution, but a
Frobenius expansion around the event horizon $r_h$:
\beq
\frac{1}{B(r)} &\!=\!& 
\gamma_1 (r\!-\!r_h) 
+\left(  \frac{3}{128\pi G\,a }-\frac{3}{128\pi G\,a\gamma_1 r_h}\!-\!\frac{2 \gamma_1}{r_h}\!+\!\frac{1}{r_h^2}  \right) (r\!-\!r_h)^2 
\!+\! \mathcal{O}(r-r_h)^3\,,\label{mc1}
\\
A(r) \!&\!=\! & 
\gamma_2\left[(r\!-\!r_h)
+\left(-\frac{1 }{128\pi G\,a  \gamma_1}\!+\!\frac{1}{128\pi G\,a \gamma_1^2 r_h}\!+\!\frac{1}{\gamma_1 r_h^2}\!-\!\frac{2}{r_h} \right) (r\!-\!r_h)^2 
\!+\! \mathcal{O}(r-r_h)^3\right]\,,\label{mc2}
\eeq
\bltx{As accomplished for the near-origin Frobenius expansion (\ref{metrica12}, \ref{metrica13}), the near-horizon case (\ref{mc1}, \ref{mc2}) has also a priori three free parameters, $a, \gamma_1, \gamma_2$, coming from the EOM (\ref{EOM2}), as 
 integration constants. 
 Implementing again the classical tests of GR \cite{Bohmer:2009yx,Casadio:2015jva}, applied to the metric (\ref{m21}) with metric coefficients (\ref{mc1}, \ref{mc2}), constrains the parameters $a, \gamma_1, \gamma_2$. In fact, 
 \beq
|\gamma_2|\lesssim |10.1\pm 12.4|\; {\rm m}^{-1}, \qquad |a\gamma_1|\lesssim |8.2\pm 10.0|\; {\rm m}^{-4}.{\rm Kg}.{\rm s}^2\qquad \label{abb1}
\eeq for the perihelion precession, whereas
\beq
|a\gamma_1^2|\lesssim |9.3\pm 37.2|\;{\rm m}^{-5}.{\rm Kg}.{\rm s}^2, \label{abb2}
\eeq for the light deflection. Besides, the radar echo delay yields the constraint
\beq
\Big|\frac{\gamma_2}{a\gamma_1}\Big|\lesssim |2.5\pm 2.7| \;{\rm m}^{3}.{\rm Kg}^{-1}.{\rm s}^{-2}.\label{abb3}
\eeq
}

One can compute the Hawking radiation flux and the black hole luminosity as well, for the metric (\ref{mc1}, \ref{mc2})
\be
\label{flux2}
L
\approxeq
\frac{\pi \,\mpl^3}{16\,l_p\,r_h}\left[1+\frac{a^2\gamma_1\gamma_2}{\pi(\gamma_1^2+2\gamma_2^2)}\right]
+\upbeta_0\,\frac{\mpl}{l_p}
\left(\frac{2\pi \,\mpl^2}{\,r_h +\frac{2a^2\gamma_1\gamma_2}{\pi(\gamma_1^2+\gamma_2^2)}}-\frac{\gamma_1\gamma_2\mpl^2r_h}{196 \left(3\gamma_1^2+4\gamma_2^2\right)\mpl}G_{\gamma}\right)
\ ,
\ee
where 
\ba
G_\gamma
&=&\frac{1}{2} \left\{\frac{a^2}{16 \gamma _2^2\mpl^2}-\frac{1}{6 \sqrt[3]{2} \gamma _2\mpl}\left(27 a^2 \gamma _1\mpl+\left[27 a^2 \gamma _1\mpl-9 a \gamma
   _2 \gamma _1\mpl^2+144 \gamma _2 \gamma _1^2\mpl^3-54 \gamma _2^3\mpl^3\right){}^2\right.\right.\nonumber\\
&&   \left.\left.-9 a \gamma _2 \gamma _1\mpl^2+2 \gamma _1^3\mpl^3+144 \gamma _2 \gamma _1^2\mpl^3-54 \gamma _2^3\mpl^3\right]^{1/3}+4 \left(3 a \gamma _2\mpl+\gamma _1^2\mpl^2+24 \gamma _2 \gamma
   _1\mpl^3\right){}^3\right.\nonumber\\&&\left.{}3\ 2^{2/3} \gamma _2\mpl \left[\sqrt{\left(27 a^2 \gamma _1\mpl-9 a \gamma _2
   \gamma _1\mpl^2+2 \gamma _1^3\mpl^3+144 \gamma _2 \gamma _1^2\mpl^3\right){}^2+4 \left(3 a \gamma _2\mpl+\gamma _1^2\mpl^2\right){}^3}\right.\right.\nonumber\\
  &&\left.\left.-9 a \gamma _2 \gamma _1\mpl^2+2 \gamma _1^3\mpl^3+144 \gamma _2 \gamma _1^2\mpl^3+\frac{\gamma _1}{3 \gamma _2}\right]^{1/3}\right\}\label{flux31}
\ea
Similarly to the analysis accomplished for the Frobenius expansion around the origin, 
the second term in Eq.~\eqref{flux2} can be negative, canceling 
the Hawking radiation out, for some critical masses,  near the Planck scale. 

After numerical routines, we derive that, in the range $0<\upbeta_0\ll 1$, the Hawking radiation flux equals zero,  \bltx{leaving black hole remnants, for
\be
M_0(\upbeta_0)\approxeq
2\upbeta_0^{-1/2}\,\mpl,\quad\qquad\text{for\;\;\; $\gamma_1=2.652,\;\; \gamma_2=3.002$\;\; $a=1.539.$}
\label{critical11}
\ee
In the range $\upbeta_0\approxeq 1$, 
the Hawking radiation flux is equal to zero when 
\be
M_0(\upbeta_0)\approxeq
4.4\,\mpl,\label{critical121}\quad\qquad\text{for\;\;\; $\gamma_1=-0.470, \;\;\gamma_2=1.381$,\;\; $a=1.227.$}
\ee
Finally, the range  $\upbeta_0\gg 1$ yields   
\be
M_0
\sim
6.1\,\mpl,\label{critical21}\quad\qquad\text{for\;\;\; $\gamma_1=1.109,\;\; \gamma_2=-1.562$,\;\; $a=2.780.$}
\ee
These particular values of the $\alpha_i$  parameters, in Eqs. (\ref{critical1} -- \ref{critical2}), are in full agreement with the 
physical bounds (\ref{abb1} -- \ref{abb3}).}
For $\upbeta=0$, when no GUP effects are taken into account, the Hawking flux does not vanish.

\bltx{Comparing the results (\ref{critical1}) -- (\ref{critical2}) and  (\ref{critical11}) -- (\ref{critical21}) with other previous results in the literature is not an easy task, since the respective black hole solutions investigated in this work are solutions of quadratic gravity. In Ref. \cite{Casadio:2017sze}, MGD black holes were studied, in a totally different context of gravity. However, also black hole remnants were shown to exist. In fact, in the range  $\beta_0\simeq 1$,  black hole remnants with mass $M_0
\sim
2.6\,\mpl$ were derived, whereas for $\beta_0\gg 1$ the critical mass reads $
M_{\rm 0}
\sim
2.2\,\mpl$. We conclude that higher-order gravity with GUP effects magnify the critical black hole masses, but keeping them of order of the Planck scale.}

\section{Concluding remarks}
\label{SIV}
The tunnelling method in the WKB approximation, in a GUP context, was here applied 
to static, spherically symmetric and asymptotically flat solutions in higher-derivative gravity with quadratic curvature terms. 
\bltx{Taking into account GUP effects, we investigated the particles tunnelling in the background of static, spherically symmetric, asymptotically flat solutions in quadratic gravity. These solutions, previously studied in Refs. \cite{Lu:2015psa,Stelle:1977ry}, contained free parameters, that have been bounded, by observational and experimental data, regarding the  the classical tests of GR in the Solar system, including the perihelion precession of Mercury, the deflection of light by the Sun
and the radar echo delay. The bounds on the parameters of the near-origin black hole solutions obtained are displayed in Eqs. (\ref{bb1}) -- (\ref{bb3}), 
whereas the constraints on the parameters of the near-horizon black hole solutions are shown in Eqs. (\ref{abb1}) -- (\ref{abb3}).
In this spacetime configurations, we showed that the corrected Hawking temperature is not only determined by the properties of the black holes, but also dependent on the mass of the emitted fermions.}

\bltx{The black hole solutions evaporation was scrutinized, in both the near-origin and near-horizon expansions.} The GUP-corrected Dirac equation was solved by the Hamilton--Jacobi method.
We showed that the Hawking flux of fermions emitted by black holes can vanish for a critical
masses~(\ref{critical1}) -- (\ref{critical2}),  respectively for fixed sets of parameters in the metric (\ref{metrica12}, \ref{metrica13}), \bltx{in the allowed range of the parameters in the black hole solutions,  obtained in Eqs. (\ref{bb1}) -- (\ref{bb3}). Besides, the Hawking radiation in the vicinity of a black hole in a higher-derivative theory
of gravity, which includes the quadratic Weyl modification
to the Einstein-Hilbert action, was also studied. The black hole evaporation that we have found is largely different from the Schwarzschild one. In fact, also in the near-horizon expansion we showed that the Hawking flux of fermions emitted by black holes can also vanish, for  critical
masses~(\ref{critical11}) -- (\ref{critical21}), in agreement with the range of  parameters in the black hole solutions  (\ref{abb1}) -- (\ref{abb3}), for fixed sets of parameters in the metric (\ref{mc1}, \ref{mc2})}. Both analyses bring the possibility of absolutely stable black hole remnants in the quadratic gravity setup. These remnants are stable also under (small) linear perturbations. The Hawking temperature of static, spherically symmetric and asymptotically flat solutions in quadratic gravity was also shown to be corrected by GUP effects.
As a perspective, one may compute the information entropy that underlies these solutions \cite{Casadio:2016aum}. 
\paragraph*{{\bf Acknowledgments}{\rm :}} 
IK  is supported by the National Council for Scientific and Technological Development  -- CNPq (Brazil) under Grant No. 155342/2018-5.  RdR~is grateful to FAPESP (Grant No.  2017/18897-8), to  CNPq (Grants No. 303390/2019-0, No.
406134/2018-9 and No. 303293/2015-2) and  to HECAP -- ICTP, Trieste, for partial financial support, and this last one also for the hospitality.

\end{document}